\documentclass [onecolumn,preprint,superscriptaddress]{revtex4-2}

\usepackage{bm}
\usepackage[colorlinks=true,linkcolor=blue,citecolor=blue]{hyperref}
\usepackage{times}
\usepackage{amsmath}
\usepackage{amssymb}
\usepackage{amsthm}
\usepackage{amsfonts}
\usepackage{enumerate}
\usepackage{latexsym}
\usepackage{ifpdf}
\usepackage{natbib}
\usepackage{psfrag}
\newcommand{\beq}{\begin{equation}}
	\newcommand{\eeq}{\end{equation}}
\usepackage{graphicx}
\usepackage{makeidx}
\usepackage{color}
\usepackage{array}
\usepackage{siunitx}
\usepackage{multirow}
\usepackage{colortbl}
\usepackage[table]{xcolor}
\usepackage{hhline}
\usepackage[table]{xcolor}
\begin{document}
	\title{Pressure-induced phase transition in pyrochlore iridates (Sm$_{1-x}$Bi$_x$)$_2$Ir$_2$O$_7$ ($x$ = 0, 0.02, and 0.10): Raman and X-ray diffraction  studies} 
	\begin{abstract}
The pyrochlore iridates, $A_2$Ir$_2$O$_7$, show a wide variety of structural, electronic, and magnetic properties controlled by the interplay of different exchange interactions, which can be tuned by external pressure. In this work, we report pressure-induced phase transitions at ambient temperature using synchrotron-based X-ray diffraction (up to $\sim$ 20 GPa) and Raman-scattering measurements (up to $\sim$ 25 GPa)  of the pyrochlore series (Sm$_{1-x}$Bi$_x$)$_2$Ir$_2$O$_7$ ($x$ = 0, 0.02, and 0.10). Our Raman and X-ray data suggest an iso-structural transition in Sm$_2$Ir$_2$O$_7$  at $P_c$ $\sim$ 11.2 GPa associated with the rearrangement of IrO$_6$ octahedra in the pyrochlore lattice. The transition pressure decreases to $\sim$  10.2 and 9 GPa for $x$ = 0.02 and 0.10, respectively. For all the samples, the linewidth of three phonons associated with Ir-O-Ir ($A_{1g}$ and $E_g$) and Ir-O ($T_{2g}^{4}$ ) vibrations show anomalous decrease up to $P_c$, due to decrease in electron-phonon interaction. 
	\end{abstract}

 \author{M Rosalin}
%\email{rosalin.zinny@email.com}
\affiliation{Department of Physics, Indian Institute of Science, Bangalore, 560012, India}
\author{K.A. Irshad}
%\email{rosalin.zinny@email.com}
\affiliation{Elettra-Sincrotrone Trieste S.C. p.A., S.S. 14, Km 163.5 in Area Science Park, Basovizza 34149, Italy}
\author{Boby Joseph}
%\email{rosalin.zinny@email.com}
\affiliation{Elettra-Sincrotrone Trieste S.C. p.A., S.S. 14, Km 163.5 in Area Science Park, Basovizza 34149, Italy}
\author{Prachi Telang}
%\email{email2@email.com}
\affiliation{Department of Physics, Indian Institute of Science Education and Research, Pune, Maharashtra, 411008, India}

\author{Surjeet Singh}
%\email{email3@email.com}
\affiliation{Department of Physics, Indian Institute of Science Education and Research, Pune, Maharashtra, 411008, India}

\author{D V S Muthu}
%\email{email3@email.com}
\affiliation{Department of Physics, Indian Institute of Science, Bangalore, 560012, India}

\author{A K Sood}
\email{asood@iisc.ac.in}
\affiliation{Department of Physics, Indian Institute of Science, Bangalore, 560012, India}

	\maketitle
	\section{Introduction}
 
Pyrochlore iridates have emerged as an interesting family of materials, providing a unique framework for studying the interplay of electronic correlations, spin-orbit coupling, and topological phenomena. These compounds, $A_2$Ir$_2$O$_7$, where $A$ is a rare earth element, have drawn particular interest due to their intricate electronic and magnetic behavior that sets the stage for the exploration of novel quantum phenomena and topological phases ~\cite{Savary2014041027, Wan2011205101, Krempa2012045124, Ahn2015115149, Kargarian2011165112}. By substituting the $A$ site successively with larger rare-earth atoms, the physical properties evolve with the $A$-site cation radius from magnetic insulating to a complex nonmagnetic metallic behavior ~\cite{Disseler2012014428, Donnerer2016037201, Klicpera2020, Graf2014012020, Nakatsuji2006087204}. In the lanthanide series, from Lu to Gd, iridates show insulating behavior ; Eu, Sm, and Nd iridates show a metal-insulator transition (MIT) with the transition temperature varying as a function of rare earth ionic radius ~\cite{Matsuhira2007043706,Zhang2017026404, Graf2014012020}. However, the end member Pr$_2$Ir$_2$O$_7$ remains metallic even at low temperatures ~\cite{Nakatsuji2006087204}. Increasing $A$ size in the $A$$_2$Ir$_2$O$_7$ series results in a broader Ir-O-Ir bond angle and shorter Ir-O bond lengths. As a result, the iridium $t_{2g}$ bandwidth widens and eventually surpasses the metallization threshold at a given $A$ ionic radius ~\cite{Zhang2017026404}.

  Application of chemical pressure i.e., doping $A^{3+}$ or Ir$^{4+}$ sites, and physical pressure,  have proven effective tools to tune the relative scale of interactions and electronic bandwidth in this class of materials. Expanding the size of the $A$ cation results in a simultaneous increase in the Ir-O-Ir bond angle and lattice parameter ~\cite{Zhang2017026404}. In contrast, under hydrostatic pressure, the Ir-O-Ir bond angle increases while the lattice parameter decreases ~\cite{Tafti2012205104}. Hence it would be interesting to explore whether the bandwidth can be tuned by applying a combination of chemical and hydrostatic pressures in a systematic manner. 
  
  Recently, it has been reported that Sm$_2$Ir$_2$O$_7$ doped with Bi (lattice constant $a$ = 10.3235 {\AA}  for Sm$_2$Ir$_2$O$_7$ and 10.3250 {\AA}  for Bi$_2$Ir$_2$O$_7$) shows anomalous lattice contraction up to 10 \% Bi doping, followed by normal lattice expansion with further Bi substitution ~\cite{telang2021}. This series of iridates provides an interesting system to see the effect of the hydrostatic pressure with systematic chemical substitution. Both Raman spectroscopy and X-ray diffraction (XRD) are powerful tools for characterizing the structural changes induced by pressure. Raman spectroscopy enables the investigation of vibrational modes, providing information about local atomic environments and structural distortions, while XRD provides information on crystallographic changes. The synergy between these techniques allows for a comprehensive analysis of the material's response to pressure. 
  
  In this study, we systematically investigate the (Sm$_{1-x}$Bi$_x$)$_2$Ir$_2$O$_7$ (SBIO) series for $x$ = 0, 0.02, and 0.10 to explore the influence of chemical tuning on the high-pressure behavior of these compounds. Our goal is to unravel how the delicate balance between electron-lattice interactions and chemical composition governs the observed structural transformations.  An iso-structural transition in Sm$_2$Ir$_2$O$_7$ at $\sim$ 11.2 GPa, marked $P_c$, is suggested by our Raman and X-ray data, which is linked to the rearrangement of IrO$_6$ octahedra in the pyrochlore lattice. The linewidths of the Ir-O-Ir bending ($A_{1g}$ and $E_g$) and Ir-O stretching ($T_{2g}^{4}$) vibrations exhibit anomalous drop with pressure up to $P_c$ for all the three samples. This decrease is caused by a decrease in the electron-phonon interaction due to the increased electronic bandwidth of these materials with pressure. The critical pressure $P_c$ decreases to $\sim$ 10.2 and 9 GPa for $x$ = 0.02 and 0.10, respectively.

	\section{Experimental details}
	  All the samples in the SBIO series ($x$ = 0, 0.02, and 0.10) were synthesized via solid-state reaction method using high purity (99.9 \%) precursors of Sm$_2$O$_3$, IrO$_2$, and Bi$_2$O$_3$. The details of sample synthesis and characterization are given in ~\cite{telang2021, Rosalin2023}.  The high-pressure XRD measurements were carried out at Xpress beamline, Elettra, in Italy, using a monochromatic XRD wavelength ($\lambda$) of 0.4957 {\AA} at room temperature. Standard CeO$_2$ powder sample was used for calibrating the sample-to-detector distance and orientation angles of the detector. The SBIO pellets ($x$ = 0, 0.02, and 0.10) were finely ground and put into membrane-type diamond-anvil cells (DACs) with a culet diameter of $\sim$ 400 $\mu$m for the XRD measurements. The pressure-transmitting medium (PTM) used for the membrane-type DAC was methanol: ethanol solution (4:1); and the applied pressure was measured using a ruby fluorescence marker ~\cite{Forman1972284,Piermarini19752774, Dewaele2004094112}. Two-dimensional (2D) diffraction patterns were captured using a Dectris Pilatus 6M detector, and subsequently, these patterns were transformed into $I$(2$\theta$) diffraction profiles by radial integration of the diffraction rings using FIT2D software ~\cite{Hammersley}. The structural analysis involved fitting and refining the raw data using a conventional Rietveld refinement procedure, facilitated by the GSAS software package ~\cite{LarsonGSAS}. High-pressure Raman scattering experiments at room temperature were carried out in backscattering geometry using a 50X objective in a Horiba LabRAM HR Evolution Spectrometer equipped with a DPSS laser source of wavelength 532 nm and $\sim$ 1.5 mW of laser power on the sample. The DAC used for Raman measurements was Mao-Bell type with two 16-facet brilliant cut diamonds with culet diameter size of $\sim$ 600 $\mu$m. The experiments were carried out on a small piece ($\sim$ 75 $\mu$m) taken from polycrystalline sintered pellets of Bi-doped Sm$_2$Ir$_2$O$_7$ samples. The sample was placed inside a stainless-steel gasket hole with a diameter of $\sim$ 150 $\mu$m.  For doped Sm$_2$Ir$_2$O$_7$, the pressure was applied using a 4:1 methanol-ethanol combination as PTM that had a freezing pressure of $\sim$  10.4 GPa ~\cite{Piermarini20035377}. However, for Sm$_2$Ir$_2$O$_7$, liquid distilled H$_2$O was used as the PTM as the methanol-ethanol solution was giving rise to a large background and poor signal-to-noise ratio.  The spectra were acquired using a thermoelectrically cooled charge-coupled device (CCD) (HORIBA Jobin Yvon, SYNCERITY 1024 $\times$ 256). The recorded Raman spectra are fitted with a sum of Lorentzian line shapes using the nonlinear least square fitting method in the Origin software package ~\cite{Deschenes2000origin} to extract phonon frequencies, linewidths, and intensities.

   \begin{figure}
		\vspace{-0pt}
		\includegraphics[width=\textwidth] {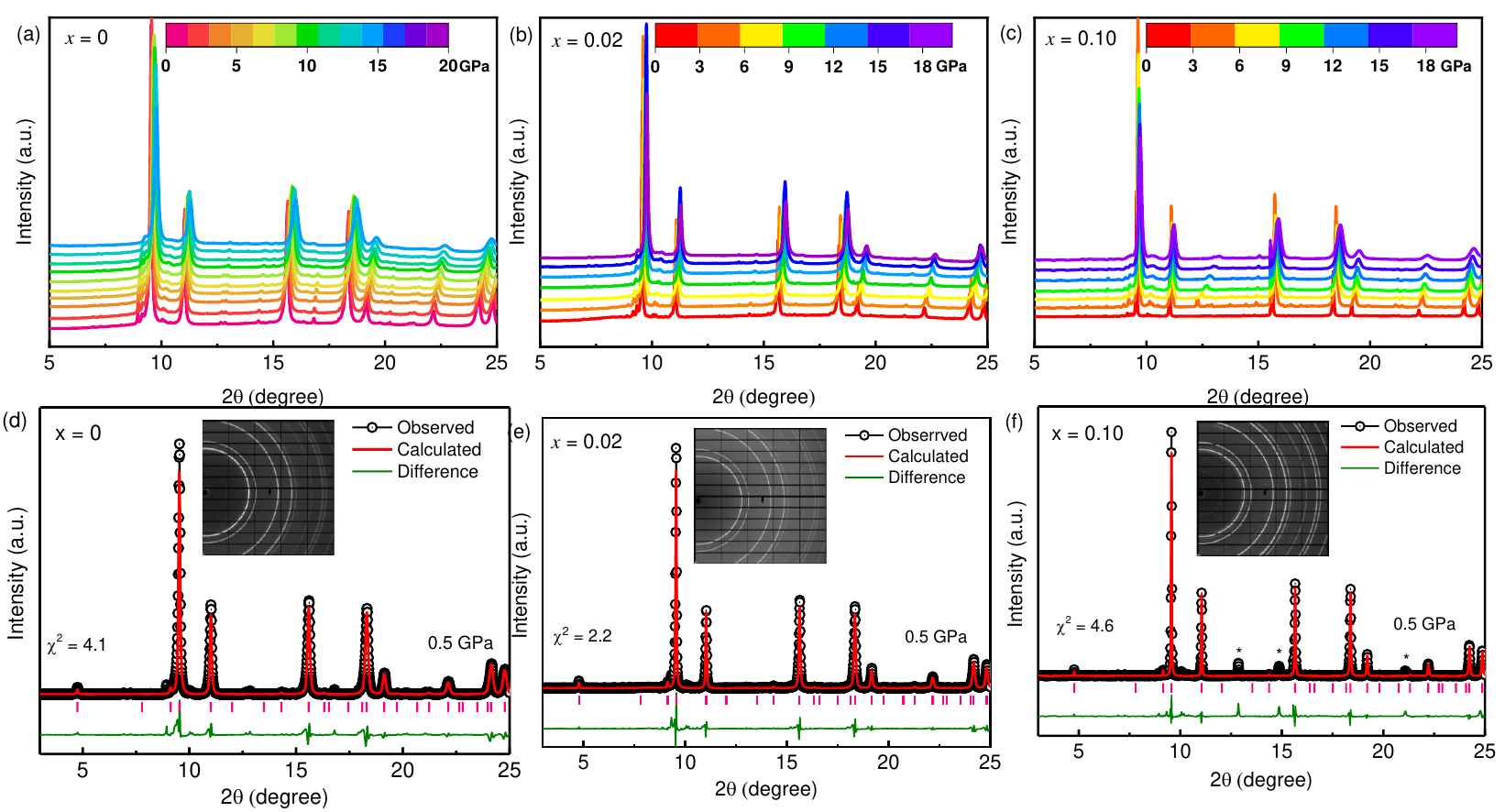}
		\caption{\label{Fig1} (a-c) Angle dispersive X-ray diffraction patterns of (Sm$_{1-x}$Bi$_x$)$_2$Ir$_2$O$_7$ (SBIO) series for $x$ = 0, 0.02, and, 0.10, respectively at selected pressures ranging from 0.2 to 20 GPa. (d-f) Rietveld refined XRD patterns at 0.5 GPa with the Debye-Scherrer ring pattern shown in the insets. Solid circles indicate experimental data and the star symbols in the (f) panel mark the Ir metal impurity peaks. Calculated patterns are drawn as red solid lines. Bragg reflection positions are indicated by vertical bars. Lower dark green curves are the difference between the observed and calculated profiles.}
	\end{figure}
	
  \section{Results and Discussions}
 \subsection{High-pressure XRD studies} The structural stability of SBIO under pressure was investigated at room temperature up to $\sim$ 20.4, 19.6, and 20.3 GPa for samples with $x$ = 0, 0.02, and 0.10, respectively. The angle dispersive powder XRD patterns of SBIO at various high pressures at 300 K are shown in Fig. 1 (a-c). These diffraction patterns show that all of the Bragg peaks shift in the direction of increasing 2$\theta$ angles as pressure increases, inferring the expected gradual contraction of the lattice. Further, the diffraction peaks significantly broaden at higher pressures due to strain-broadening. However, the absence of any new Bragg peak or of existing peaks splitting indicates absence of structural phase transition under pressure. Figure 1 (d-f) shows the Rietveld fitted diffraction profiles for ambient pressure cubic pyrochlore phase having space group Fd$\overline{3}$m ($z$ = 8). The lattice parameters, atomic positions, line
shape, and background were all refined in this process and the goodness-of-fit parameter ($\chi^2$) given in Fig. 1 indicates that the quality of the refinements remains good throughout the SBIO series.  In Fig. 2 (a-c), the volume of the unit cell (estimated using the lattice parameter value from Rietveld refinement of SBIO series) is plotted against pressure which shows the cubic lattice volume decreases with increasing pressure. The $P$-$V$ data are fitted by a third-order Birch-Murnaghan (BM) equation of state (EOS) ~\cite{Birch19864949} given by,

\begin{equation}
\begin{aligned}
P= & \frac{3 B}{2}\left[\left(\frac{V_0}{V}\right)^{\frac{7}{3}}-\left(\frac{V_0}{V}\right)^{\frac{5}{3}}\right] \\
& \times\left\{\left[1+\frac{3}{4}\left(B^{\prime}-4\right)\right]\left[\left(\frac{V_0}{V}\right)^{\frac{2}{3}}\right]\right\}
\end{aligned}
\end{equation}
where $B$ represents the bulk modulus or the modulus of incompressibility at ambient pressure, $B'$ represents the first derivative of the bulk modulus with respect to pressure, and $V_0$ represents the unit cell volume at ambient pressure. The extracted values from the fit are $B$ = 198.4 $\pm$ 4.0 GPa and $B'$ = 7.8 $\pm$ 0.8 for $x$ = 0, which compares well with the previously reported value for Eu$_2$Ir$_2$O$_7$ ~\cite{Clancy2016024408, Thomas2023Highpressure} and recent preprint on  Sm$_2$Ir$_2$O$_7$ ~\cite{stasko2024robustness}. The fit parameters for doped samples are, $B$ = 214.7 $\pm$ 6.3 GPa and $B'$ = 8.5 $\pm$ 2.4 for $x$ = 0.02, and $B$ = 250.5 $\pm$ 9.6 GPa and $B'$ = 18.5  $\pm$ 4.0 GPa for $x$ = 0.10. The value of $B'$ becomes noticeably large in the case of $x$ = 0.10, indicating that doped SBIO ($x$ = 0.10) stiffens rapidly under pressure. 
\begin{figure}
		\vspace{-0pt}
		\includegraphics[width=\textwidth] {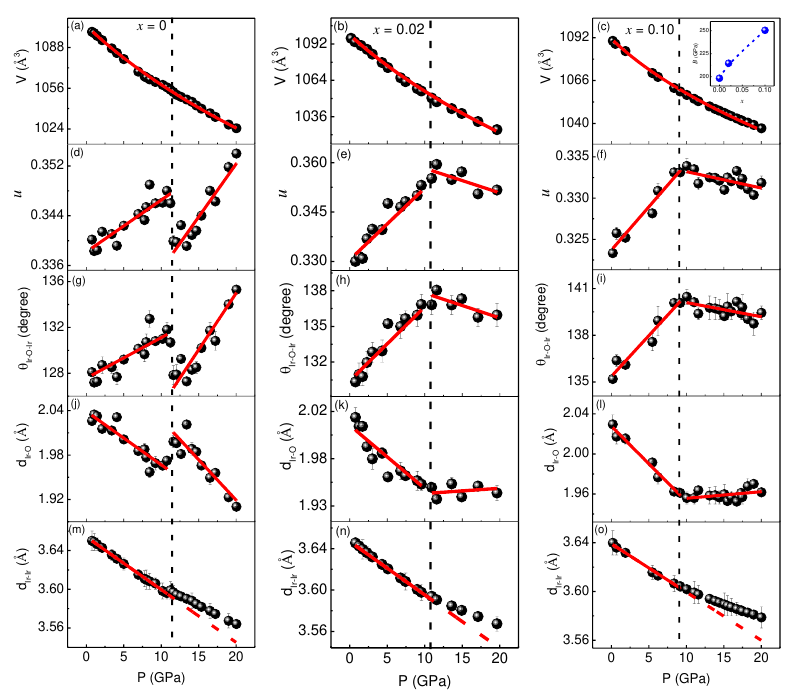}
		\caption{\label{Fig2} (a-c) Pressure dependence of lattice volume in the pyrochlore phase of SBIO ($x$ = 0, 0.02, and 0.10, respectively), fitted (solid red line) $P$-$V$ diagram using a third-order Birch–Murnaghan equation of state. Inset in the rightmost top panel show the variation of the bulk modulus ($B$) with doping concentration $x$.(d-o) The pressure dependence of the Rietveld refinement parameters, Oxygen coordinate ($u$), Ir-O-Ir bond angle ($\theta_{Ir-O-Ir}$), Ir-Ir (d$_{Ir-Ir}$), and Ir-O bond distance (d$_{Ir-0}$),  respectively for SBIO ($x$ = 0, 0.02, and 0.10). The vertical dashed lines in each panel indicate $P_c$ and the linear fit below $P_c$ is extrapolated above $P_c$ by dashed red lines in panels m,n, and o.}
  \end{figure}
It is clear from Fig. 2 that the $P$-$V$ plot for all the samples can be fitted over the entire pressure window upto  $\sim$ 20 GPa.  However, the signatures of a transition at $P_c$, not reported so far, are clearly seen in the pressure dependence of the unit cell parameters derived from the Rietveld refinement plotted in Fig. 2 (d-o). In the pyrochlore structure, each Ir-site cation is coordinated to six O (48f) ions, forming an IrO$_6$ octahedron which has a variable $u$-coordinate, the only free parameter in the pyrochlore structure besides the lattice parameter, $a$. The perfect octahedral symmetry is achieved for $u$ = 0.3125. The pyrochlore stability field lies within the interval of 0.3125 $\leq$ $u$ $\leq$ 0.375, and the IrO$_6$ octahedral exhibits a trigonal compression that increases with $u$ ~\cite{Telang2018235118,SUBRAMANIAN198355,Clancy2016024408}. The evolution of $u$ as a function of pressure in Fig. 2 (d) shows a finite increase with a jump near $P_c$, suggesting that the IrO$_6$ octahedra become increasingly more distorted at higher pressures with a sudden change in the degree of distortion occurring at $P_c$, likely due to an iso-structural transition implying a rearrangement of the IrO$_6$ octahedra. The increase in $u$ in Sm$_2$Ir$_2$O$_7$ with pressure is similar to Eu$_2$Ir$_2$O$_7$ ~\cite{Clancy2016024408}, but in the recent high-pressure preprint on $A_2$Ir$_2$O$_7$ ($A$ = Pr, Sm, Dy-Lu), $u$ remains constant over increasing pressure ~\cite{stasko2024robustness}.   Since the pressure-dependent XRD patterns do not reveal any new diffraction peak or splitting of lines, it is inferred that the structural deformation at $\sim$ $P_c$ is a local distortion. It is possible that as the pressure increases, due to vacancies at the 8a sites, the Ir$^{4+}$ ions adjust their local coordinates, causing other atoms in the lattice to relocate, retaining the cubic symmetry and affecting oxygen positions ($u$) in the IrO$_6$ octahedra as reflected in the abrupt change in $u$ near at $P_c$ ~\cite{Saha2009134112}.   The corresponding Ir-O-Ir bond angle (Fig. 2 (g)) and Ir-O bond length (Fig. 2 (j)) also show discontinuities near $P_c$. The pressure coefficient of Ir-Ir bond length (Fig. 2 (m)) also shows a clear change at $P_c$. The inset of Fig. 2 (c) shows the increase of the bulk modulus with the doping concentration $x$. Increasing values of bulk modulus with increasing $x$ is expected, following the anomalous lattice contraction in SBIO from $x$ = 0 to 0.10. It may be noted that all the four lattice characteristics for $x$ = 0.02 and 0.10 samples also change at $P$ $\sim$ 10.2 and 9 GPa, respectively in Fig. 2 (e-f, h-i, k-l, n-o). The decrease in $P_c$ with $x$ can be attributed to the anomalous lattice contraction in SBIO. 
 \begin{figure}
		\vspace{-0pt}
		\includegraphics[width=0.85\textwidth] {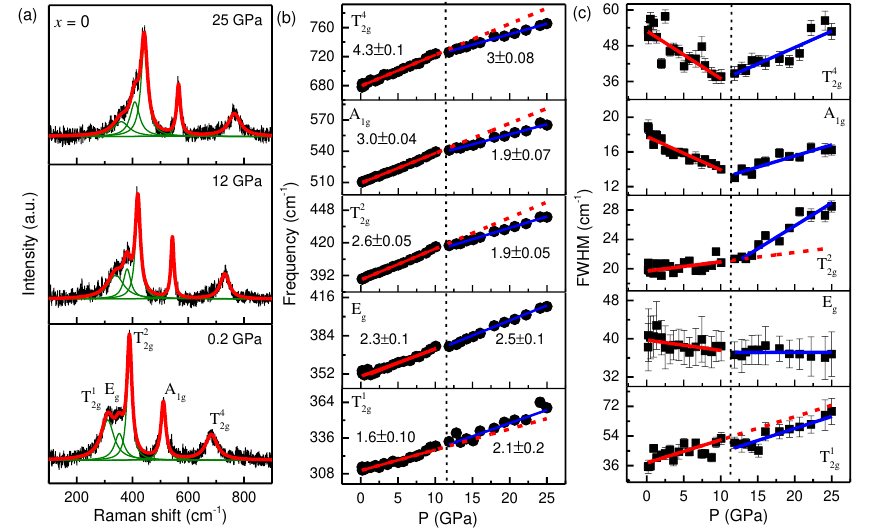}
		\caption{\label{Fig3}(a) Fitted Raman spectra of Sm$_{2}$Ir$_{2}$O$_{7}$ at some representative pressures. Dark green curves are individual fit to the phonon modes, and solid red curves are cumulative fit to the spectra. (b-c) The pressure dependence of frequencies and linewidths, respectively, of various phonon modes up to 25 GPa. Solid red and blue lines are linear fit to the data below and above $P_c$, respectively. The vertical dashed lines indicate $P_c$ and the linear fit below $P_c$ is extrapolated above $P_c$ by dashed red lines in (b) and (c).}
	\end{figure}
 \subsection{High-pressure Raman studies on Sm$_{2}$Ir$_{2}$O$_{7}$}

 Based on group theory, the pyrochlore structure (space group Fd$\overline{3}$m) has zone-center optical, $\Gamma_{op}$ = $A_{1g}$ + $E_g$ + 4$T_{2g}$ + 7$T_{1u}$  and acoustic, $\Gamma_{ac}$ = $T_{1u}$ phonon modes ~\cite{Ueda2019115157}. The Raman active phonons are $A_{1g}$, $E_g$, and 4$T_{2g}$, and the rest seven are IR active. The Raman modes are assigned with irreducible representations, as shown in Fig. 3 (a), based on prior reports on pyrochlore iridates ~\cite{Ueda2019115157,KumarHarish202313178}. The mode assignments are as follows : $A_{1g}$ and $E_g$ modes involve Ir-O-Ir bending, $T_{2g}^{1}$ and $T_{2g}^{2}$ modes associated with Sm-O stretching, $T_{2g}^{3}$ mode due to the vibration of O’ ions surrounded by eight Sm ions and $T_{2g}^{4}$ mode related to Ir-O stretching vibration ~\cite{Ueda2019115157}. As the cation sites of Ir$^{4+}$ and Sm$^{3+}$ have center of inversion, the Raman active modes involve only the vibrations of oxygen atoms. Figure S1 shows the pressure evolution of the Raman modes at some representative pressures.  Raman spectrum at the lowest pressure inside the DAC for Sm$_{2}$Ir$_{2}$O$_{7}$ is plotted in the bottommost panel of Fig. 3 (a), which shows five phonon modes. The position and relative intensities of these bands agree well with the literature ~\cite{Ueda2019115157, KumarHarish202313178}. $T_{2g}^{3}$ is a weak mode and could not be detected inside the DAC. Raman spectra at an intermediate and highest pressure are shown in the middle and topmost panel in Fig. 3 (a). The same number of phonon modes are observed with varying pressure which confirms the preservation of crystal symmetry with pressure, in agreement with the XRD results. The individual phonon modes were fitted with Lorentzian line shape as shown in Fig. 3 (a), and the extracted phonon frequencies and linewidths are plotted against pressure in Figs. 3 (b) and (c). The straight lines represent the fitting of mode frequencies using the linear equation, $\omega_P=\omega_0+\left(\frac{d \omega}{d P}\right) P$, where $\omega_0$ is the mode frequency at ambient pressure. All the phonon modes demonstrate expected hardening with increasing pressure due to the compression of the unit cell, resulting in an increase in the stiffness constant ($K$) associated with the vibrational mode. However, we observe an alteration of the pressure-coefficients ($d\omega/dP$) of the phonon mode frequencies across $P_c$ $\sim$ 11.2 GPa, supporting the presence of an iso-structural transition as inferred by the XRD results.  It should be noted that we have performed the high-pressure experiment with distilled water as the PTM, which starts freezing above 1 GPa, making it a quasi-hydrostatic pressure medium.  Therefore, the anomalies near $P_c$ are unlikely to be triggered by the non-hydrostatic nature of the medium. Further, our XRD data using methanol: ethanol as PTM also exhibit anomalies near $P_c$, indicating that the transition is intrinsic to the system. A previous high-pressure study on Gd$_{2}$Ti$_{2}$O$_{7}$ has also shown that the pressure dependence of the Raman modes is seen to be similar in both methanol-ethanol mixture and distilled water as PTM ~\cite{Saha2006064109}. The two high-pressure iso-structural phases are associated with a positive slope value for all the phonon modes, as depicted in Fig. 3 (b). To quantify the phonon frequency changes under pressure we have estimated the Gr\"{u}neisen parameter ($\gamma$) given by,
\begin{equation}
\gamma=\frac{B}{\omega_0} \frac{\mathrm{d} \omega}{\mathrm{d} P}
\end{equation} 
below and above $P_c$ for all the phonon modes. The values are listed in Table I. The significant changes in  $\gamma$ suggest a subtle structural deformation, as seen in pressure-dependent XRD measurements (see Fig. 2).

 \begin{table}[ht]
\centering
\begin{tabular}{|c|c|c|}
\hline 
Modes & $P < P_c$ & $P > P_c$ \\
\hline 
$T_{2g}^{1}$ & 1.04 & 1.35 \\ 
\hline 
$E_g$ & 1.28  & 1.42 \\
\hline 
$T_{2g}^{2}$ & 1.35  & 0.95  \\ 
\hline
$A_{1g}$ & 1.12 & 0.72  \\ 
\hline
$T_{2g}^{4}$ & 1.25  & 0.85  \\ 
\hline
\end{tabular}
\caption{ Gr\"{u}neisen parameter ($\gamma$) for various phonon modes below and above $P_c$ for $x$ = 0}
\end{table}
 
    To understand the frequency variations in Sm$_{2}$Ir$_{2}$O$_{7}$, we compared the effect of physical and chemical pressure, similar to our earlier high-pressure work on Eu$_{2}$Ir$_{2}$O$_{7}$ ~\cite{Ueda2019115157}. As an effect of $A$- site cation radius, the lattice volume decreases by $\sim$ 32 ${\AA}^3$, coming from Pr$_{2}$Ir$_{2}$O$_{7}$ to Sm$_{2}$Ir$_{2}$O$_{7}$ in the pyrochlore series ~\cite{Takatsu2014235110}. The difference in phonon frequencies for the $A_{1g}$ mode in Sm$_{2}$Ir$_{2}$O$_{7}$ compared to Pr$_{2}$Ir$_{2}$O$_{7}$ is $\sim$ 2 cm$^{-1}$ ~\cite{Ueda2019115157}. Our $P$-$V$ data (Fig. 2 (a)) indicates that a volume reduction of 32 ${\AA}^3$ equates to a pressure of $\sim$ 7 GPa, causing the $A_{1g}$ mode frequency to harden by $\sim$ 18 cm$^{-1}$.  The significant difference in frequency shifts for the $A_{1g}$ phonon due to physical and chemical pressures cannot be explained only by the quasiharmonic effect, indicating that the $A_{1g}$ phonon is also renormalized by an underlying electron-phonon interaction. It can be seen from Fig. 2 that the Ir-Ir and Ir-O bond distances systematically decrease up to $P_c$, and thus increasing electronic bandwidth in a metal by enhancing the hybridization of the Ir-5$d$ and O-2$p$ orbitals, and direct hopping between the Ir sites ~\cite{Zhang2017026404, Tafti2012205104}. Similarly, an increase in the Ir-O-Ir bond angle would favor electrons hopping between two Ir atoms via the oxygen atom and contribute to the increasing bandwidth. Thus, the large frequency variation of the $A_{1g}$ mode in Sm$_{2}$Ir$_{2}$O$_{7}$ can be correlated with the increase in the electronic bandwidth leading to significant changes in the electron-phonon interaction ~\cite{Debernardi2001213}.

Figure 3 (c) depicts the pressure evolution of linewidths for the five strongest Raman phonons, showing an unexpected drop in the linewidth of the Ir-O-Ir bending ($A_{1g}$ and $E_g$) as well as Ir-O stretching vibrations $T_{2g}^{4}$  upto $P_c$. The solid lines are linear fit to $\Gamma=\Gamma_0+\left(\frac{d \Gamma}{d P}\right) P$. It can be seen that $\frac{d \Gamma}{d P}$ is negative for $A_{1g}$, $E_g$, and $T_{2g}^{4}$ modes. The values of $\frac{d \Gamma}{d P}$ are given in table II. This observation is similar to our recent high-pressure Raman study on Eu$_{2}$Ir$_{2}$O$_{7}$, where an equivalent linewidth drop was seen for the $A_{1g}$ phonon ~\cite{Thomas2023Highpressure}. One possible explanation for the anomalous decrease in the linewidth of these modes is a decrease in the electron-phonon interaction due to pressure. Since this anomaly is observed in phonons related to Ir-O and Ir-O-Ir vibrations, it can be correlated with the observed systematic trend of the Ir–Ir and Ir-O bond distance as well as the Ir-O-Ir bond angle up to $P_c$ which increases the electronic bandwidth, thereby decreasing the electron-phonon interaction.  According to the Holstein model, the electronic-phonon interaction strength in a metal, $\lambda=2 g^2 / W \omega$ (where $\omega$ is the phonon frequency, $t$ is the nearest-neighbor hopping integral, $W$ = 8$t$ is the electronic bandwidth, and $g$ is the coupling constant), will vary inversely with its electronic bandwidth ~\cite{Dee2020-lz}. As seen in Fig. 2, $\theta_{Ir-O-Ir}$ increases with pressure, increasing the electronic bandwidth and thus decreasing $\lambda$ and the observed linewidth drop. The linewidth of other $T_{2g}^{1}$ and $T_{2g}^{2}$ phonons increases with increasing pressure which can be attributed to the corresponding change in the two-phonon density of states, which affects the anharmonic interactions ~\cite{Serrano2004014301}.
  \begin{figure}
		\vspace{-0pt}
		\includegraphics[width=0.65\textwidth] {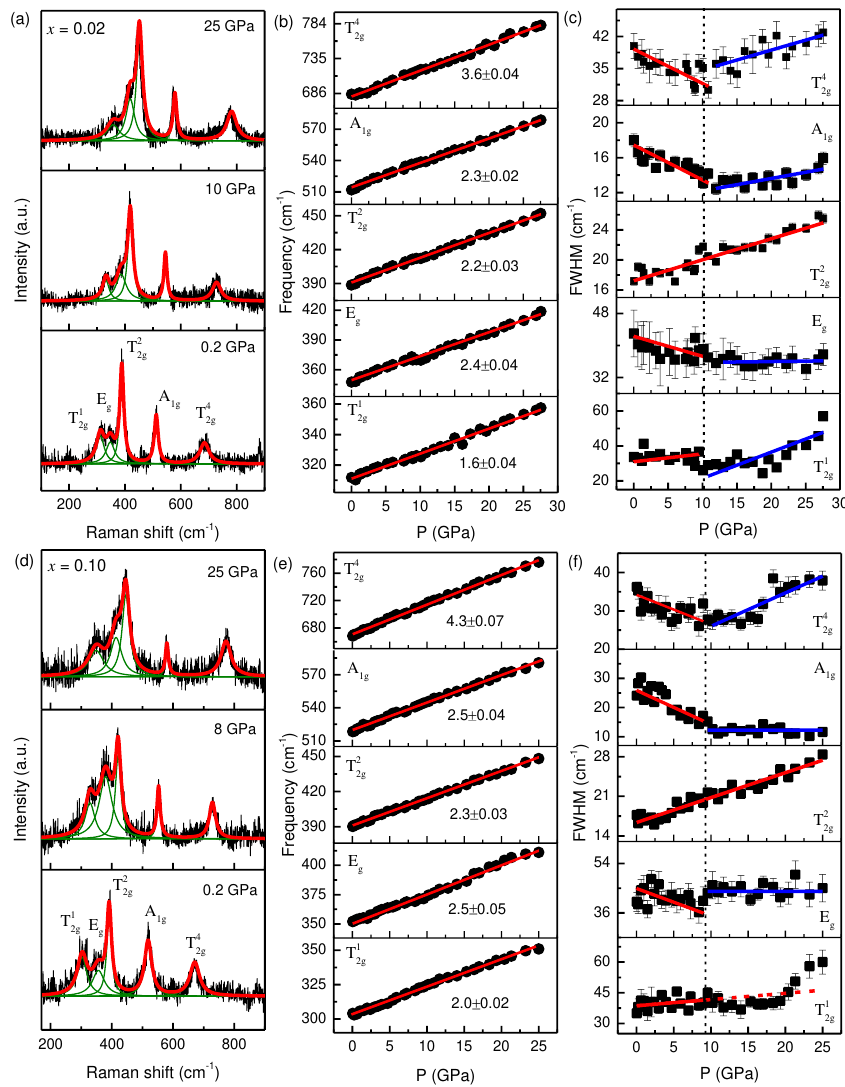}
		\caption{\label{Fig4} (a,d) Fitted Raman spectra of SBIO ($x$ = 0.02 and 0.10, respectively) at some representative pressures.  ((b,c) and (e,f) The pressure dependence of frequencies and linewidths of various phonon modes for $x$ = 0.02 and 0.10, respectively. Symbols and lines have the same meaning as Fig. 3.}
	\end{figure}

\begin{table}[ht]
\centering
\setlength{\arrayrulewidth}{1pt} % Adjust thickness of vertical lines
\arrayrulecolor[rgb]{0,0,0} % Set line color to black
\begin{tabular}{|>{\columncolor[RGB]{244, 240, 240}}c|>{\columncolor[RGB]{255, 255, 204}}c|>{\columncolor[RGB]{255, 255, 204}}c|>{\columncolor[RGB]{230, 242, 255}}c|>{\columncolor[RGB]{230, 242, 255}}c|>{\columncolor[RGB]{255,204,255}}c|>{\columncolor[RGB]{255,204,255}}c|}
\hline 
\cellcolor[gray]{0.9} & \multicolumn{2}{|>{\columncolor[RGB]{255, 255, 204}}c|}{ $x = 0$ } & \multicolumn{2}{|>{\columncolor[RGB]{230, 242, 255}}c|}{ $x = 0.02$ } & \multicolumn{2}{|>{\columncolor[RGB]{255,204,255}}c|}{ $x = 0.10$ } \\
%Column 1 & Column 2 & Column 3 & Column 4 & Column 5 & Column 6 & Column 7 \\
\hhline{|~|------|}
%\hhline{2-7}
\multirow{-2}{*}[-5pt]{\cellcolor[gray]{0.9}Modes} & $P < P_c$ & $P > P_c$ & $P < P_c$ & $P > P_c$ & $P < P_c$ & $P > P_c$ \\
\hline 
$T_{2g}^{1}$ & 1.4 $\pm$ 0.3 & 1.5 $\pm$ 0.2 & 0.45 $\pm$ 0.2 & 1.5 $\pm$ 0.4 & 0.31 $\pm$ 0.3 & - \\
\hline 
$E_g$ & -0.2 $\pm$ 0.08 & 0 & -0.5 $\pm$ 0.2 & 0.02 $\pm$ 0.07 & -1 $\pm$ 0.04 & 0\\
\hline 
$T_{2g}^{2}$ & 0.13 $\pm$ 0.05 & 0.62 $\pm$ 0.05 & 0.3 $\pm$  0.02 & 0.3 $\pm$  0.02 & 0.44 $\pm$ 0.02 & 0.44 $\pm$ 0.02\\
\hline
$A_{1g}$ & -0.4 $\pm$ 0.04 & 0.25 $\pm$ 0.03 & -0.4 $\pm$ 0.03 & 0.14 $\pm$ 0.03 & -1.2 $\pm$ 0.3 & 0\\
\hline
$T_{2g}^{4}$ & -1.6 $\pm$ 0.3 & 1.1 $\pm$ 0.17 & -0.7 $\pm$ 0.1 & 0.43 $\pm$ 0.1 & -0.75 $\pm$ 0.2 & 0.87 $\pm$ 0.14\\
\hline
\end{tabular}
\caption{ $\frac{d \Gamma}{d P}$ for various phonon modes below and above $P_c$}
\end{table}

% \begin{table}[ht]
% \centering
% \setlength{\arrayrulewidth}{1pt} % Adjust thickness of vertical lines
% \setlength{\arrayrulewidth}{1pt} % Adjust thickness of horizontal lines
% \arrayrulecolor[rgb]{0,0,0} % Set line color to black
% \begin{tabular}{|>{\columncolor[RGB]{244, 240, 240}}c|>{\columncolor[RGB]{255, 255, 204}}c|>{\columncolor[RGB]{255, 255, 204}}c|>{\columncolor[RGB]{230, 242, 255}}c|>{\columncolor[RGB]{230, 242, 255}}c|>{\columncolor[RGB]{255,204,255}}c|>{\columncolor[RGB]{255,204,255}}c|}
% \hline 
% \cellcolor[gray]{0.9} & \multicolumn{2}{|>{\columncolor[RGB]{255, 255, 204}}c|}{ $x = 0$ } & \multicolumn{2}{|>{\columncolor[RGB]{230, 242, 255}}c|}{ $x = 0.02$ } & \multicolumn{2}{|>{\columncolor[RGB]{255,204,255}}c|}{ $x = 0.10$ } \\
% \cline{2-7}
% \multirow{-2}{*}{\cellcolor[gray]{0.9}Modes} & $P < P_c$ & $P > P_c$ & $P < P_c$ & $P > P_c$ & $P < P_c$ & $P > P_c$ \\
% \hline 
% $T_{2g}^{1}$ & 1.4 $\pm$ 0.3 & 1.5 $\pm$ 0.2 & 0.45 $\pm$ 0.2 & 1.5 $\pm$ 0.4 & 0.31 $\pm$ 0.3 & - \\
% \hline 
% $E_g$ & -0.2 $\pm$ 0.08 & 0 & -0.5 $\pm$ 0.2 & 0.02 $\pm$ 0.07 & -1 $\pm$ 0.04 & 0\\
% \hline 
% $T_{2g}^{2}$ & 0.13 $\pm$ 0.05 & 0.62 $\pm$ 0.05 & 0.3 $\pm$  0.02 & 0.3 $\pm$  0.02 & 0.44 $\pm$ 0.02 & 0.44 $\pm$ 0.02\\
% \hline
% $A_{1g}$ & -0.4 $\pm$ 0.04 & 0.25 $\pm$ 0.03 & -0.4 $\pm$ 0.03 & 0.14 $\pm$ 0.03 & -1.2 $\pm$ 0.3 & 0\\
% \hline
% $T_{2g}^{4}$ & -1.6 $\pm$ 0.3 & 1.1 $\pm$ 0.17 & -0.7 $\pm$ 0.1 & 0.43 $\pm$ 0.1 & -0.75 $\pm$ 0.2 & 0.87 $\pm$ 0.14\\
% \hline
% \end{tabular}
% \caption{ $\frac{d \Gamma}{d P}$ for various phonon modes below and above $P_c$}
% \end{table}

 \subsection{High-pressure Raman studies on SBIO ($x$ = 0.02 and 0.10)}
 The pressure evolution of the Raman spectra of $x$ = 0.02 and 0.10 are given in Figs. S2-S3, showing the same number of phonon modes up to the highest measured pressure. Figure 4 (a) and (d) shows fitted Raman spectra for SBIO ($x$ = 0.02 and 0.10, respectively) at a few representative pressures, marking  $E_g$, $A_{1g}$, and three $T_{2g}$ phonons. The corresponding phonon frequencies are presented in panels (b) and (e) of this figure, with pressure coefficients listed alongside the straight-line fits to the data. For all phonons, the frequency versus pressure plot fits with a single positive slope ($\frac{d \omega}{d P}$ ) over the whole recorded pressure range. The lack of slope change at $P_c$ in doped Sm$_{2}$Ir$_{2}$O$_{7}$ can be understood by the fact that the IrO$_6$ octahedra are less deformed at ambient pressure due to a smaller value of the oxygen coordinate $u$, and the iso-structural rearrangement is smaller as compared to the undoped sample. The linewidth of the $A_{1g}$, $E_g$, and  $T_{2g}^{4}$ modes show anomalous decrease up to $P_c$ ($\sim$ 10.2 and 9 GPa for x = 0.02 and 0.10, respectively) as shown in panels (c) and (f) of Fig. 4, similar to the $x$ = 0 sample. Figure 5 (a) depicts the evolution of the pressure coefficient of the linewidth ($\frac{d \Gamma}{d P}$) as a function of doping concentration ($x$) which demonstrates that the $T_{2g}^{4}$ phonon for $x$ = 0 has a greater linewidth decrease than the $A_{1g}$ and $E_g$ phonons, whose amplitude decreases with Bi doping. The  $\frac{d \Gamma}{d P}$ of the $A_{1g}$ and $E_g$ phonons is similar for $x$ = 0 and 0.02 but increases significantly for $x$ = 0.10 sample.  These variations in the linewidths of $A_{1g}$, $E_g$, and  $T_{2g}^{4}$ modes infer that the Ir-O-Ir bending and Ir-O bond stretching vibrations are influenced differently by electron-phonon interaction. Our Raman data clearly reveal that the crossover pressure $P_c$ for the doped sample decreases gradually with $x$ similar to (Eu$_{1-x}$Bi$_x$)$_2$Ir$_2$O$_7$ ~\cite{Thomas2023Highpressure}. 
\begin{figure}
		\vspace{-0pt}
		\includegraphics[width=\textwidth] {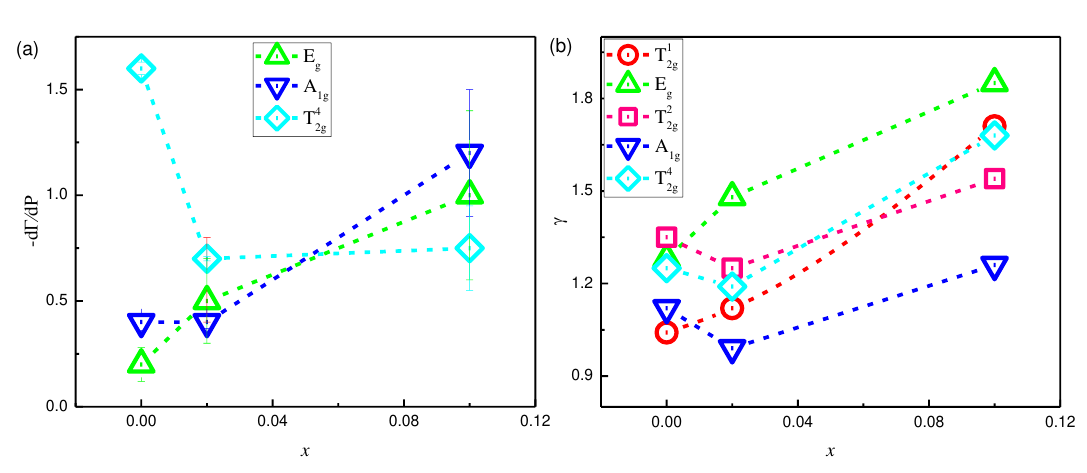}
		\caption{\label{Fig5} (a,b) Evolution of $\frac{d \Gamma}{d P}$ and Grüneisen parameter below $P_c$ ($\gamma$ ), respectively of various Raman phonons with doping concentration $x$. Dashed lines are guides to the eye.}
	\end{figure}
Figure 5 (b) shows the evolution of Gr\"{u}neisen parameter ($\gamma$) (below $P_c$) with doping concentration $x$ for five phonon modes, showing a monotonic increase, as in (Eu$_{1-x}$Bi$_x$)$_2$Ir$_2$O$_7$ ~\cite{Thomas2023Highpressure}, reflecting the increase of bulk modulus with $x$ (see inset of Fig. 2 (c)).
  
\section{Conclusions}
 In summary, we have investigated pressure-induced structural changes in the (Sm$_{1-x}$Bi$_x$)$_2$Ir$_2$O$_7$ ($x$ = 0, 0.02, and 0.10) series using synchrotron-based X-ray diffraction and Raman-scattering measurements up to $\sim$ 20 and 25 GPa, respectively. Our Raman and X-ray results suggest an iso-structural phase transition associated with the rearrangement of IrO$_6$ octahedra in the pyrochlore lattice in Sm$_2$Ir$_2$O$_7$, at $\sim$ 11.2 GPa. The transition pressure decreases to $\sim$ 10.2 and 9 GPa for $x$ = 0.02 and 0.10, respectively due to an anomalous lattice contaction in SBIO series. The linewidths of the Ir-O-Ir bending ($A_{1g}$ and $E_g$) and Ir-O stretching ($T_{2g}^{4}$) vibrations exhibit anomalous drop with increasing pressure up to $P_c$ for all the samples, attributed to a decrease in electron-phonon interaction. The integration of Raman spectroscopy and XRD measurements in this study not only enhances the reliability of our observations but also allows for a comprehensive understanding of the structural changes occurring in SBIO under high pressure. Our findings contribute to the fundamental understanding of pyrochlore iridates and hold potential implications for the design of materials with tailored properties and novel quantum states.

\section{Acknowledgement}
 AKS thanks the Department of Science and Technology, Government of India, for financial support under the National Science Chair Professorship. The authors acknowledge the financial support by the Department of Science and Technology (DST) of the Government of India for the high-pressure XRD measurements at the Xpress beamline of Elettra Sincrotrone, Trieste.

\bibliographystyle{aipnum4-2}
\bibliography{reference,reference}

\begin{figure}
		\vspace{-0pt}
		\includegraphics[width=0.75\linewidth] {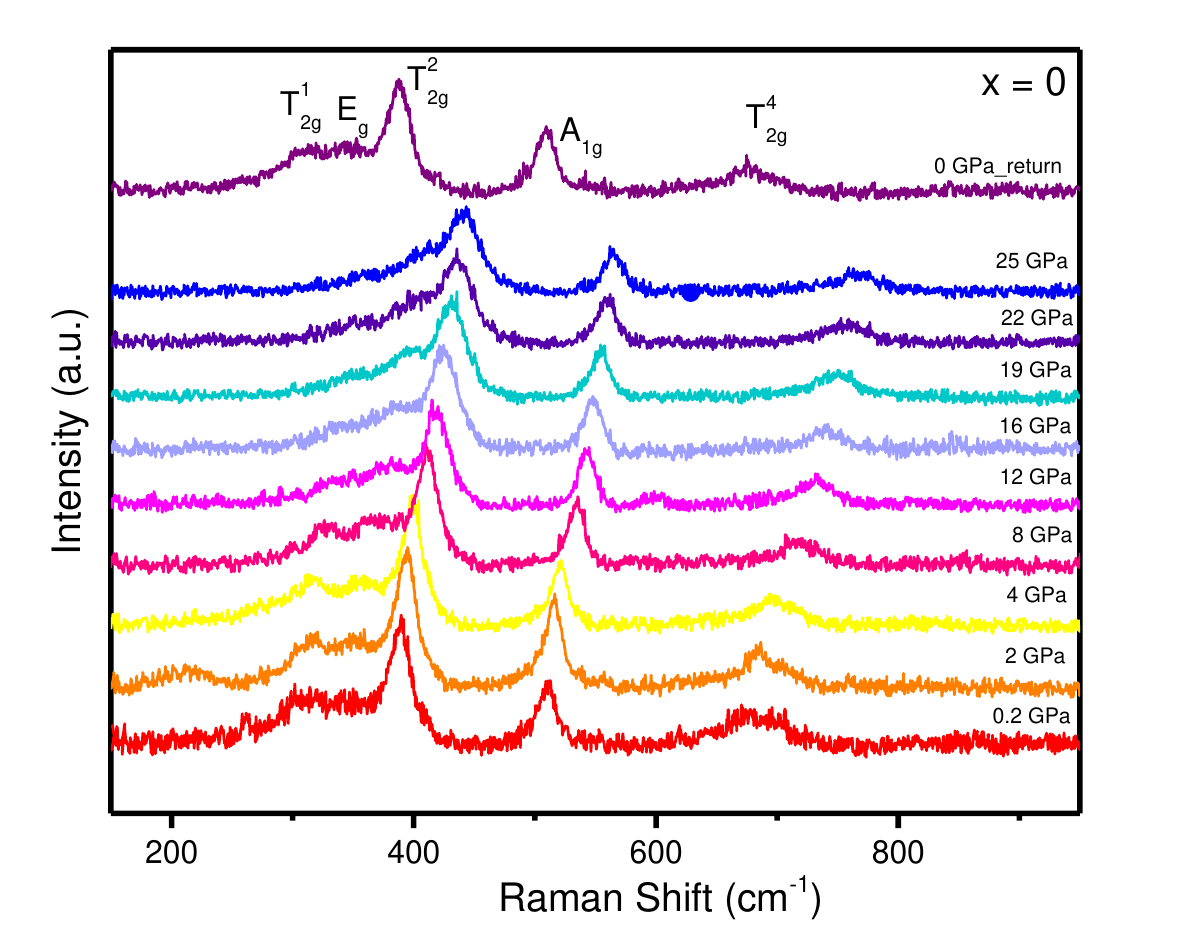}
		\caption{ [Fig.S1 : Pressure evolution of the Raman spectra for $x$ = 0.]}
  \label{FigS1}
	\end{figure}
  \begin{figure}
		\vspace{-0pt}
		\includegraphics[width=0.75\linewidth] {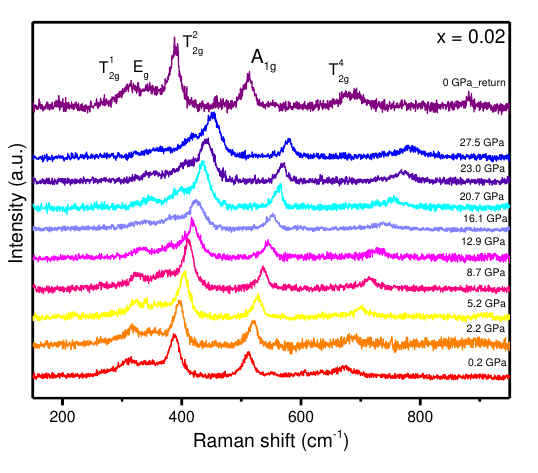}
		\caption{[Fig.S2 : Pressure evolution of the Raman spectra for $x$ = 0.02.] }
  \label{FigS4}
	\end{figure}
\begin{figure}
		\vspace{-0pt}
		\includegraphics[width=0.75\linewidth] {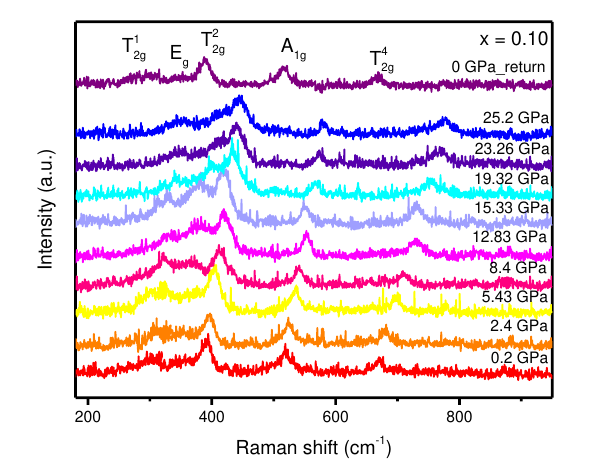}
		\caption{[Fig.S3 : Pressure evolution of the Raman spectra for $x$ = 0.10.]}
  \label{FigS4}
	\end{figure}
%aipnum4-2.bst 2019-01-14 (MD) hand-edited version of apsrev4-1.bst

\end{document}